\documentclass[prd,preprint,floatfix,nofootinbib]{revtex4}
\usepackage{epsfig}
\usepackage{graphicx}
\usepackage{pstricks}

\def\lsim{\mathrel {\vcenter {\baselineskip 0pt \kern 0pt
    \hbox{$<$} \kern 0pt \hbox{$\sim$} }}}
\def\gsim{\mathrel {\vcenter {\baselineskip 0pt \kern 0pt
    \hbox{$>$} \kern 0pt \hbox{$\sim$} }}}
\def\slashchar#1{\setbox0=\hbox{$#1$}           
 \dimen0=\wd0                                 
  \setbox1=\hbox{/} \dimen1=\wd1               
\ifdim\dimen0>\dimen1                        
  \rlap{\hbox to \dimen0{\hfil/\hfil}}      
  #1                                        
  \else                                        
 \rlap{\hbox to \dimen1{\hfil$#1$\hfil}}   
   /                                         
  \fi}                                         %
\def\cpto{\mathrel {\vcenter {\baselineskip 0pt \kern 0pt
    \hbox{$CP$} \kern 0pt \hbox{$\longrightarrow$} }}}
\def\cptof{\mathrel {\vcenter {\baselineskip 0pt \kern 0pt
    \hbox{$~CP$} \kern 0pt \hbox{$\longleftrightarrow$} }}}

\begin{document}

\baselineskip=15pt

\preprint{}

\title{B decays with $\tau$-leptons in non-universal left-right models}

\author{Xiao-Gang He}
\email{hexg@phys.ntu.edu.tw}
\affiliation{INPAC, SLKPPC and Department of Physics, Shanghai Jiao Tong University,  Shanghai, China\\
Department of Physics, National Tsing Hua University, and National Center for Theoretical Sciences, Hsinchu, Taiwan\\
CTS, CASTS and Department of Physics,  National Taiwan University, Taipei, Taiwann}

\author{German Valencia}
\email{valencia@iastate.edu}
\affiliation{Department of Physics and Astronomy, Iowa State University, Ames, IA 50011, USA}

\date{\today $\vphantom{\bigg|_{\bigg|}^|}$}

\date{\today}

\vskip 1cm
\begin{abstract}

Recent measurements of semileptonic $B$-meson decays into $\tau$-leptons are somewhat higher than expected in the standard model. Although the deviations are less than 3$\sigma$, they suggest the possibility of new physics affecting primarily the $\tau$-lepton. In this paper we examine these results within the context of non-universal left-right models. We find that strong constraints from $b\to s \gamma$ on $W-W^\prime$ mixing lead to a prediction of approximately equal enhancements for the $B\to D \tau \nu$ and the $B\to D^\star \tau \nu$ modes. The model predicts approximately the same enhancement for the inclusive semileptonic rate $\bar{B} \to X_c \tau\nu$ as well as for the leptonic decay $B_c^-\to \tau^- \nu$. An enhancement in the leptonic decay $B^-\to \tau^-\nu$ is also possible but is not uniquely correlated with the other modes. For this explanation to be viable, the mass of the non-universal $W^\prime$ would be below 1~TeV,  within LHC reach.

\end{abstract}

\pacs{PACS numbers: }

\maketitle

\section{Introduction}

Recent measurements of $B$ decay modes involving $\tau$-leptons have shown hints of deviations from the standard model (SM). In particular, for the semileptonic $b\to c \tau \nu$ modes, for which 
BaBar has reported recently that \cite{Lees:2012xj}
\begin{eqnarray}
R(D) &=& \frac{B(\bar{B}\to D\tau^-\bar{\nu}_\tau)}{B(\bar{B}\to D\ell^-\bar{\nu}_\ell)}\, = \, 0.440\pm 0.072 \nonumber \\
R(D^\star) &=& \frac{B(\bar{B}\to D^\star \tau^-\bar{\nu}_\tau)}{B(\bar{B}\to D^\star\ell^-\bar{\nu}_\ell)}\, = \, 0.332\pm 0.030 
\label{babarsemi}
\end{eqnarray}
The corresponding numbers from Belle are \cite{Bozek:2010xy}
\begin{eqnarray}
R(D) &=& \frac{B(\bar{B}\to D\tau^-\bar{\nu}_\tau)}{B(\bar{B}\to D\ell^-\bar{\nu}_\ell)}\, = \, 0.35\pm 0.11 \nonumber \\
R(D^\star) &=& \frac{B(\bar{B}\to D^\star \tau^-\bar{\nu}_\tau)}{B(\bar{B}\to D^\star\ell^-\bar{\nu}_\ell)}\, = \, 0.43 \pm 0.08 
\label{bellesemi}
\end{eqnarray}
Both the BaBar and Belle results are a bit high compared to the SM expectation  \cite{Kamenik:2008tj,Fajfer:2012vx},
\begin{eqnarray}
R(D) &=& 0.297 \pm 0.017 \nonumber \\
R(D^\star) &=& 0.252 \pm 0.003
\end{eqnarray}
At the same time the discrepancy between theory and experiment for the leptonic $B\to \tau \nu$ rate appears to be getting smaller, with the latest Belle result with hadronic tags being $0.72^{+0.27}_{-0.25}\pm0.11)\times 10^{-4}$ \cite{Adachi:2012mm}. Below we quote Belle's combination of this result with their result based on semileptonic tags \cite{Adachi:2012mm}, as well as BaBar's number \cite{Lees:2012ju} and a recent SM prediction by the CKMfitter group  \cite{ckmfitter}.
\begin{eqnarray}
B(B\to \tau \nu)_{Belle} &=& (0.96 \pm 0.26)\times 10^{-4} \,  \nonumber \\
B(B\to \tau \nu)_{BaBar} &=& (1.79 \pm 0.48)\times 10^{-4}\,  \nonumber \\
B(B\to \tau \nu)_{SM} &=& (0.719^{+0.115}_{-0.076} )\times 10^{-4} .\,
\label{fbtau}
\end{eqnarray}

The apparent discrepancy in the semileptonic modes has inspired their examination  within two Higgs doublet models.  The data disfavors type II models \cite{Tanaka:2010se} where it is not possible to simultaneously accommodate the enhancements in $R(D)$ and $R(D^\star)$ \cite{Lees:2012xj}. Type III 2HDM have been also studied and are consistent with the data \cite{Crivellin:2012ye}. 2HDM models are also discussed in this context in Ref.~\cite{Celis:2012dk}. The results have also been discussed in the context of R-parity violation \cite{Deshpande:2012rr} and more generally with the aim of making further predictions \cite{efflags}. They have also inspired new calculations of the form factors \cite{Faustov:2012nk}.

In this paper we consider another new physics possibility to address this enhancement. In the context of non-universal left-right (LR) models \cite{heval}, these modes receive additional contributions from the right-handed charged currents that are suppressed or absent for modes involving the first two generation leptons. The model contains one light right-handed neutrino that would  ultimately be responsible for the enhancements in the semileptonic $B$-meson decays by adding the channel $\tau\nu_R$ to the possible final states.

\section{Non-universal LR Model}

The model we consider is a variation of LR models \cite{lrmodel}, and we have detailed it previously \cite{heval}. Here we review the salient features relevant for our discussion. 
The gauge group is the usual $SU(3)_C\times SU(2)_L\times SU(2)_R\times U(1)_{B-L}$. With this gauge group there are two charged gauge bosons, $W_L$ and $W_R$ for $SU(2)_L$ and $SU(2)_R$, respectively. In general they mix with each other and the mass eigenstates $W$ (light) and $W'$ (heavy) are a mixture of $W_{L}$ and $W_R$. This mixing can be parameterized as
\begin{eqnarray}
W_L &=& \cos\xi_W W - \sin \xi_W W'\;,\nonumber\\
W_R &=& \sin\xi_W W +\cos\xi_W W'\;.
\end{eqnarray}
The mass splitting between $W$ and $W'$ in this model is achieved by introducing the Higgs multiplets $H_L: (1, 2, 1)(-1)$ and $H_R: (1, 1, 2)(-1)$, and by letting
their vacuum expectation values (vev) $v_L$ and $v_R$ be different. In our notation for the quantum numbers of the multiplets,  the three numbers in the first bracket are the representations under $SU(3)_C$, $SU(2)_L$ and $SU(2)_R$ respectively. The number in the second bracket is for $B-L$. The $W_L-W_R$ mixing can be induced by a bi-doublet scalar $\phi: (1, 2, 2)(0)$ in which  both of the diagonal fields in $\phi$ have non-zero vev $v_{1,2}$.

The non-universal version of the L-R model we consider here is the one  proposed in Ref.~\cite{heval}. In this model, only the third generation interacts with $W_R$, and the first two generations have SM like interactions. The first two generations (indicated by the superscript) thus transform under the gauge group as
\begin{eqnarray}
&&Q_L^{1,2}: (3, 2, 1)(1/3)\;,\;\;\;\;U^{1,2}_R: (3,1,1)(4/3)\;,\;\;\;\;D_R^{1,2}: (3,1,1)(-2/3)\;,\nonumber\\
&&L_L^{1,2}: (1, 2, 1)(-1)\;,\;\;\;\;E^{1,2}_R: (1,1,1)(-2)\;.
\end{eqnarray}

The third generation is different, as its right handed particles also transform under $SU(2)_R$. The transformation of the third generation under the gauge group is given by
\begin{eqnarray}
&&Q_L^{3}: (3, 2, 1)(1/3)\;,\;\;\;\;Q^3_R: (3,1,2)(1/3)\;,\nonumber\\
&&L_L^{3}: (1, 2, 1)(-1)\;,\;\;\;\;L_R^3: (1,1,2)(-1)\;.
\end{eqnarray}
This model is anomaly free. It accommodates an enhanced interaction between third generation fermions that becomes of electroweak strength for process that involve a fermion pair from the first two generations and one from the third generation. Its phenomenology has been studied in detail, with the most stringent bounds on the $Z^\prime$ arising from LEP2 \cite{heval} and most recently from the study of $pp\to \tau^+\tau^-$ at the LHC \footnote{Analysis of this data within our model is not finished but we expect a constraint on the $Z^\prime$ mass around 1~TeV, of the same order as those quoted by CMS for other $Z^\prime$ models.} \cite{Chatrchyan:2012hd}. We are not aware of any LHC searches for non-universal $W^\prime$ bosons so far, but this paper would be an additional motivation to conduct them.

In the mass eigenstate basis the quark-gauge-boson interactions are 
given by,
\begin{eqnarray}
{\cal L}_W&=& - {g_L\over \sqrt{2}} \bar U_L \gamma^\mu V_{KM} D_L
(\cos\xi_W W^{+}_\mu - \sin\xi_W W^{'+}_\mu)\nonumber\\
&&-{g_R\over \sqrt{2}}
\bar U_{R} \gamma^\mu V_{R} D_{R}
(\sin\xi_W W^{+}_\mu + \cos\xi_W W^{'+}_{\mu}) ~+~{\rm h.~c.},
\label{cccoup}
\end{eqnarray}
where $U = (u,\;\;c,\;\;t)$, $D = (d,\;\;s,\;\;b)$, $V_{KM}$ is
the Kobayashi-Maskawa mixing matrix and $V_R = (V_{Rij})=(V^{u*}_{Rti}V^{d}_{Rbj})$ with $V^{u,d}_{Rij}$ the unitary matrices
which rotate the right handed quarks $u_{Ri}$ and $d_{Ri}$ from the weak eigenstate basis 
to the mass eigenstate basis. The model contains FCNC and their phenomenology has also been studied in detail, it is summarized in Ref.~\cite{He:2009ie}.

To have an acceptable neutrino mass spectrum and mixing, additional fields need to be introduced.
For this purpose we introduce the  Higgs triplet representations, $\Delta_L: (1, 3, 1)(2)$ and $\Delta_R: (1, 1, 3)(2)$. For our model to be interesting for semileptonic $B$ decays, it must have a light right-handed neutrino. This implies that the vevs  $v^L_\Delta$ and $v^R_\Delta$  for the two $\Delta_L$ and $\Delta_R$ fields need to be small. In particular $v^R_\Delta$ is not as large as in the usual Left-Right seesaw models. The neutrinos obtain Majorana masses from $v^{L,R}_\Delta$ in addition to the  Dirac masses induced by the non-zero vev of $\phi$. The resulting neutrino mass eigenstates $(\nu_{L}^m, (\nu_{R3}^m)^c)$ are related
to the weak eigenstates $(\nu_{L}, \nu_{R3}^c)$ by

\begin{eqnarray}
\left (\begin{array}{l} \nu_L\\\nu_{R3}^c \end{array}\right ) =
\left ( \begin{array}{ll}
U_L&U_{RL}\\U_{LR}&U_R \end{array}\right ) \left ( \begin{array}{l} \nu^m_L\\ (\nu^m_{R3})^c\end{array} \right)\;.
\end{eqnarray}
Note that $\nu_L$ and $\nu^m_L$ have three components.

Since only the third generation has a right-handed neutrino, $U_{L} = (U_{Lij})$ is a $3\times 3$ matrix determined by the Majorana mass term for left-handed neutrinos. $U_{RL}=(U_{RLi3})$ and $U_{LR}=(U_{LR3i})$ are $3\times1$ and $1\times 3$ matrices, respectively,  determined by the Dirac mass terms 
for neutrinos, and $U_R=(U_{R33})$ is a $1\times 1$ matrix from the Majorana mass term for the right-handed neutrino.

As already mentioned, we consider the scenario where the new right-handed neutrino is very light (even though it has a Majorana mass), as this is the only case relevant for $B$ decays.  If the third generation neutrino is mostly Dirac type, the elements in $U_{LR, RL}$ can be large, of order one. In this case $\nu^c_{L3}$ is dominated by the weak eigenstate $\nu_{R3}$. If the right-handed neutrino is Majorana, $U_R$ can be large. If the right-handed neutrino has a mass negligible compared with the $\tau$ mass, these two cases give the same results for our purposes.

In general the charged leptons will also mix with each other, and their weak and mass eigenstates are related by: $\ell_{L,R} = V^\ell_{L,R}\ell^m_{L,R}$
with $\ell = (e, \mu, \tau)^T$. Here $V^\ell_{L,R}$ are the unitary matrices which diagonalize the charged lepton mass matrix.
In the lepton mass eigenstate basis (dropping the superscript ``m'' on the fields), we have

\begin{eqnarray}
{\cal L}_W &=& - {g_L\over \sqrt{2}} (\bar \nu_L \gamma^\mu U^{\ell \dagger}\ell_L 
+ \bar \nu_{R3}^c \gamma^\mu U^{\ell *}_{RLj3}\ell_{Lj})
W^{+}_\mu (\cos\xi_W W^{+}_\mu - \sin\xi_W W^{'+}_\mu) \nonumber \\
&-&{g_R\over \sqrt{2}}
(\bar \nu^c_{Li} \gamma^\mu  U^\ell_{LRij} \ell_{Rj} + \bar \nu_{R3} \gamma^\mu U^\ell_{R3j}\ell_{Rj})
(\sin\xi_W W^{+}_\mu + \cos\xi_W W^{'+}_{\mu})  ~+~{\rm h.~c.},
\label{cccouplep}
\end{eqnarray}
where
\begin{eqnarray}
U^{\ell\dagger} = U^{ \dagger}_L V^\ell_L\;,\;\;
U^{\ell *}_{RLj3} = (U_{RLi3}^{*}V^\ell_{Lij})\;,\;\;
U^\ell_{LRij} = U_{LR3i} V^\ell_{R3j}\;,\;\;
U^\ell_{R3j}=U_{R33} V^\ell_{R3j}\;.
\end{eqnarray}
$U^\ell$ is approximately the PMNS matrix.

In $B$ meson decays  the final neutrino flavor is not identified so it must be summed.  For the processes involving left- and right-handed charged leptons, neglecting neutrino masses compared with the charged lepton masses, the final decay rates into a charged lepton $\ell_j$ are proportional to 
\begin{eqnarray}
\mbox{For $\ell_{Lj}$}: &&\sum_i |U^\ell_{ij}|^2 +|U^\ell_{RLj3}|^2  = \sum_i |U^*_{Lli}V^\ell_{Llj}|^2 +|U^*_{RLl3}V^\ell_{Llj}|^2\nonumber\\
&&=(\sum_i |U^*_{Lli}|^2 +|U^*_{RLl3}|^2)|V^\ell_{Llj}|^2= \sum_l |V^\ell_{Llj}|^2 = 1\;,\nonumber\\
\mbox{For $\ell_{Rj}$}: &&\sum_i |U^\ell_{LRij}|^2 +|U^\ell_{R3j}|^2  = \sum_i |U_{LR3i}V^\ell_{R3j}|^2 +|U_{R33}V^\ell_{R3j}|^2\nonumber\\
&&=(\sum_i |U_{LR3i}|^2 +|U_{R33}|^2)|V^\ell_{R3j}|^2 =|V^\ell_{R3j}|^2\;.
\end{eqnarray}
In obtaining these results we used the unitarity of $U$: $\sum_i |U^*_{Lli}|^2 +|U^*_{RLl3}|^2=1$, 
$\sum_i |U_{LR3i}|^2 +|U_{R33}|^2=1$ and the unitarity of $V_{L,R}^\ell$.

All this implies that within this model, the leptonic and semileptonic decay modes of the $B$ meson involving $\tau$-leptons  will turn out to depend on the following combinations of parameters (two for each case $q=u,c$ and for each lepton $\ell$):
\begin{eqnarray}
F^q_{W^\prime} &=& \left( 1+
\left(\frac{g_RM_W}{g_L M_{W^\prime}}\right)^4\frac{|V^\ell_{R3\ell}|^2|V_{Rqb}|^2}{{|V_{qb}|^2}}\right) \nonumber \\
F^q_{{\rm Mix}} &=& \xi_W\frac{g_R}{g_L}\frac{{\rm Re}\left(V_{qb}^\star V_{Rqb} \right)}{{|V_{qb}|^2}}\left(1-\left(\frac{M_W}{M_{W^\prime}}\right)^2\right)\left(1+\left(\frac{g_RM_W}{g_L M_{W^\prime}}\right)^2|V^\ell_{R3\ell}|^2\right).
\label{newfs}
\end{eqnarray}
The first term arises from direct $W^\prime$ contributions and the second term appears through $W-W^\prime$ mixing contributions. In the next section we summarize our results in terms of the combinations of Eq.~\ref{newfs}.

\section{Leptonic and semileptonic $B$-meson decay modes into $\tau$-leptons}

\subsection{Inclusive semileptonic $B$ decay}

The inclusive semileptonic decay rate is easily calculated to first order in the $W-W^\prime$ mixing parameter $\xi_W$ using the free quark model. The differential distribution with respect to the dimensionless momentum transfer $z=M_{\ell\nu}^2/m_b^2$, $M_{\ell\nu}^2=(p_b-p_c)^2=(p_\ell+p_\nu)^2$, can be written in terms of the combinations defined in Eq.~\ref{newfs} as

\begin{eqnarray}
\frac{d\Gamma\bigl(b\to q \ell^-\bar\nu_{\tau}\bigr)}{dz}&=&\frac{G_{\rm F}^2\,m_b^5}{192\pi^3}\,|V_{qb}|^2\,\sqrt{\lambda(1,r_q^2,z)}\ \left[F^q_{W^\prime}\ \frac{2(z-r_\ell^2)^2}{z^3}\right.\nonumber\\
&\cdot &\left(z(r_q^4+r_q^2(z-2)-2z^2+z+1)-r_\ell^2(z^2+z(1+r_q^2)-2(1-r_q^2)^2) \right)\nonumber\\
&-& \left.24\ F^q_{\rm Mix}\ r_q
\frac{(z-r_\ell^2)^2}{z}\right]
\label{incldis}
\end{eqnarray}
in terms of the usual kinematic function
\begin{eqnarray}
\lambda(x,y,z)=x^2+y^2+z^2-2(xy+xz+yz),
\label{kinlam}
\end{eqnarray}
and of the ratios $r_q=m_q/m_b$ and $r_\ell=m_\ell/m_b$. 
For electrons and muons, where the lepton mass can be neglected, this can be easily integrated to obtain the total decay rate,
\begin{eqnarray}
\Gamma\bigl(b\to q \ell^-\bar\nu^{}\bigr)&=&
\frac{G_{\rm F}^2\,m_b^5}{192\pi^3}\, |V_{qb}|^2\,\left[
F^q_{W^\prime}\ F\left(\frac{m_q^{2}}{m_b^{2}}\right)  + \frac{1}{2} \ F^q_{\rm Mix}\ G\left(\frac{m_q^{2}}{m_b^{2}}\right) \right]
\label{semiinc}
\end{eqnarray}
where $F(r)$ and $G(r)$ are the known kinematic functions given by
\begin{eqnarray}
F(r) &=& 1-8r+8r^3-r^4-12r^2\,\ln r \nonumber \\
G(r)&=&-8\sqrt{r}\left(1+9r-9r^2-r^3+6r(1+r)\,\ln r\right).
\end{eqnarray}
Notice that the term proportional to $W-W^\prime$ mixing behaves as an anomalous $cbW$ coupling. In the $M_{W^\prime}>>M_W$ limit it corresponds to 
\begin{eqnarray}
\kappa^R_{cb} &=& \frac{g_R}{g_L}\frac{V^R_{cb}}{V_{cb}}\xi_W
\end{eqnarray}
in the notation of Ref.~\cite{He:2009hz}.\footnote{There is an error in Eq. 18 of Ref.~\cite{He:2009hz}. Instead of 2G it should say G/2 and this leads to $-0.015 \leq {\rm~Re}(\kappa_{cb}^R) \leq +0.088$.}

It follows from Eq.~\ref{semiinc} that, in the limit where  $W-W^\prime$ mixing is negligible, the CKM angles extracted from inclusive $b$ decay are modified as follows
\begin{eqnarray}
V_{qb}^{\rm eff} &=& V_{qb} \left( 1+
\left(\frac{g_RM_W}{g_L M_{W^\prime}}\right)^4\frac{|V^\ell_{R3\ell}|^2|V^R_{qb}|^2}{{|V_{qb}|^2}}\right)^{\frac{1}{2}} 
\end{eqnarray}
and would be different in decays into electrons or muons according to the mixing angle $V^\ell_{R3\ell}$. This would need to be taken into account in a global fit using our model. Since there is no evidence of a difference in $V_{qb}$ as extracted from electron or muon modes we will take 
$V^\ell_{R3e}=V^\ell_{R3\mu}$. Also,  we are interested in new physics that affects mostly the $\tau$-lepton, so we will assume for now that $V^\ell_{R3e}=V^\ell_{R3\mu} \approx 0$, so that  $V^\ell_{R3\tau} \approx 1$.

For decays into $\tau$-leptons, $\ell=\tau$, the kinematics is more complicated but Eq.~\ref{incldis} can be integrated to give
\begin{eqnarray}
\Gamma\bigl(b\to q \tau^-\bar\nu_{\tau}\bigr)&=&
\frac{G_{\rm F}^2\,m_b^5}{192\pi^3} \, |V_{qb}|^2 \, \left[F^q_{W^\prime}\ 
\tilde{F}(r_q^2,r_\tau^2)
- 4\ F^q_{\rm Mix}\ \tilde{G}(r_q^2,r_\tau^2)\right]
\end{eqnarray}
in terms of the kinematic form factors
\begin{eqnarray}
\tilde{F}(x,y) &=& \sqrt{\lambda}\left(1-7(x+y)-7(x^2+y^2)+x^3+y^3+xy(12-7(x+y))\right) \nonumber\\
&+& 12\left[x^2{\rm Log}\left(\frac{(1+x-y+\sqrt{\lambda})^2}{4x}\right)+y^2{\rm Log}\left(\frac{(1+y-x+\sqrt{\lambda})^2}{4y}\right)
\right. \nonumber \\
&& \left. -x^2y^2{\rm Log}\left(\frac{(1-x-y+\sqrt{\lambda})^2}{4xy}\right)\right] \nonumber \\
\tilde{G}(x,y) &=&  \sqrt{x}\left[ \sqrt{\lambda}\left(1+x^2+10x-5y(1+x)-2y^2\right) \right.\nonumber \\
&+&\left. 6x(x+(1-y)^2){\rm Log}\left(\frac{(1-y)(1-y-\sqrt{\lambda})-x(1+y)}{x(1-x+y-\sqrt{\lambda})}\right)\right.\nonumber \\
&-&\left.6y^2(1-x){\rm Log}\left(\frac{(1-x+y-\sqrt{\lambda})^2}{4y}\right)\right] 
\label{inclusivetau}
\end{eqnarray}
where we have used $\lambda$ as a shorthand notation for $\lambda(1,x,y)$ as defined in Eq.~\ref{kinlam}. 
The $g_R\to 0$ limit of Eq.~\ref{inclusivetau} agrees with the known SM result  \cite{semitaumass,Falk:1994gw}. Corrections to the free-quark decay model, of order ${\cal O}(1/m_b^2)$ have been calculated in the literature for the SM case \cite{Falk:1994gw} and can be added to our results for detailed comparisons with experiment.

For our later estimates it will be useful to have the numerical results for the kinematic form factors. Taking $m_c=1.25$ and $m_b=4.2$ we obtain $\tilde{F}=0.069$ and $\tilde{G}= 0.014$. 

\subsection{Leptonic decay $B^\pm \to \ell^\pm \nu_\ell$}

The rate for the leptonic decay $B_q^\pm \to \tau^\pm \nu_\tau$ ($q=u,c$) is given in terms of the $B_q$-meson decay constant $f_{B_q}$ defined by
\begin{eqnarray}
<0|\bar{q}\gamma^\mu \gamma_5 b|B_q(p)> &=& -i f_{B_q} p^\mu
\end{eqnarray}
as
\begin{eqnarray}
\Gamma(B^\pm \to \ell^\pm \nu_\ell) &=& \frac{G_F^2}{8\pi} f_{B_q}^2m_\ell^2 m_{B_q}\left(1-\frac{m_\ell^2}{m_{B_q}^2}\right)^2\ |V_{qb}|^2 \left(F^q_{W^\prime}\ 
- 2\ F^q_{\rm Mix}
  \right)
\label{leptonic}
\end{eqnarray}

For the leptonic decay only the axial-vector current contributes. In the limit of no $W-W^\prime$ mixing, Eq.~\ref{leptonic} reflects that there is no interference between the $W$ and $W^\prime$ contributions due to the different Dirac structure in the lepton leg.

\subsection{Semileptonic decay $B\to D \tau \nu$}

To calculate the exclusive semileptonic decay rate for $B\to D \tau \nu$ we use the standard parameterization for the form factors with $q=p-p^\prime$
\begin{eqnarray}
<D(p^\prime)|\bar{c}\gamma^\mu\ b|\bar{B}(p)> &=& f_+(q^2) (p+p^\prime)^\mu\ +\ f_-(q^2) (p-p^\prime)^\mu.
\label{dff}
\end{eqnarray}

In terms of these form factors and the kinematic function of Eq.~\ref{kinlam}, the differential decay rate can be written as
\begin{eqnarray}
\frac{d\Gamma(B\to D \tau \nu)}{dz} &=&  \frac{G_{\rm F}^2\,m_B^5}{192\pi^3}
\frac{\sqrt{\lambda(1,r_D^2,z)}(z-r_\ell^2)^2}{2z^3}\left[f_+^2 \left(2\lambda(1,r_D^2,z) z+r_\ell^2\left(\lambda(1,r_D^2,z)+3(1-r_D^2)\right)\right) \right. \nonumber \\
&+& \left.f_-^2(3r_\ell^2z^2)-f_+f_-(6(r_D^2-1)r_\ell^2z\right] \ |V_{qb}|^2\ \left(F^q_{W^\prime}\ 
+ 2\ F^q_{\rm Mix}\right)
\label{diffford}  
\end{eqnarray}
where  $z=M_{\ell \nu}^2/M_B^2=q^2/M_B^2$,  $r_D=M_D/M_B$ and $r_\ell$ was defined above.

It is interesting that this expression, Eq.~\ref{diffford}, is proportional to the SM result. This occurs because the decay is only sensitive to the vector form factor which has the same sign for both $W$ and $W^\prime$ contributions. This result allows us to write the decay rate in this case as
\begin{eqnarray}
\Gamma(B\to D \tau \nu) &=& \Gamma(B\to D \tau \nu)_{SM} \ \left(F^q_{W^\prime}\ 
+ 2\ F^q_{\rm Mix}\right)
\end{eqnarray}
and no further knowledge of the form factors Eq.~\ref{dff} is needed in order to compare with the experimental results.

\subsection{Semileptonic decay $B\to D^\star \tau \nu$}

To calculate the exclusive semileptonic decay rate for $B\to D^\star \tau \nu$ we use one of the standard form factor parameterizations \cite{Manohar:2000dt}, modified so that all the form factors are dimensionless:
\begin{eqnarray}
<D^\star(p^\prime)|\bar{c}\gamma^\mu\gamma_5\ b|\bar{B}(p)> &=&
-i M_B f(z)   \epsilon^{\star \mu} -i \frac{\epsilon^\star\cdot p}{M_B}  \left(
a_+(z)\ (p+p^\prime)^\mu + a_-(z)\ (p-p^\prime)^\mu \right)\nonumber \\
<D^\star(p^\prime)|\bar{c}\gamma^\mu\ b|\bar{B}(p)> &=&\frac{g(z)}{M_B} \epsilon^{\mu\nu\alpha\beta}\ \epsilon^\star_\nu\ (p+p^\prime)_\alpha (p-p^\prime)_\beta .
 \label{dstarff}
\end{eqnarray}
In terms of the dimensionless variables $z=M_{\ell \nu}^2/M_B^2=q^2/M_B^2$, $r_{D^\star}=M_{D^\star}/M_B$ and $r_\ell=M_\ell/M_B$ we obtain
\begin{eqnarray}
\frac{d\Gamma(B\to D^\star \tau \nu)}{dz} &=&  \frac{G_{\rm F}^2\,m_B^{ 5}}{192\pi^3}\ |V_{qb}|^2
\left[ F^q_{W^\prime}\ 
\left( F_V + F_A\right) 
+ 2  \ F^q_{\rm Mix} \left( F_V - F_A\right)   \right] 
\end{eqnarray}
where the two kinematic form factors are now given by
\begin{eqnarray}
F_V& =& g(z)^2\ \frac{ \lambda^{3/2} }{z^2}\ (2z+r_\ell^2)\ (r_\ell^2-z)^2    \nonumber\\
F_A&=& \frac{ \lambda^{1/2} (r_\ell^2-z)^2}{8 r_{D^\star}^2 z^3}\ \left[ 2\left(z\left(\lambda + 12 r_{D^\star}^2 z\right)+2r_\ell^2 \left(\lambda+3r_{D^\star}^2z\right)\right)
\ f(z)^2 \ \right. \nonumber \\
&+& 3 \lambda r_\ell^2 z^2 \ a_-(z)^2 \ 
 + \lambda \left( 2\lambda z+ r_\ell^2\left( \lambda+3(r_{D^\star}^2-1)^2\right) \right) \ a_+(z)^2 \nonumber \\
 &+& 6 \lambda r_\ell^2 z \left( (r_{D^\star}^2-1) \ a_+(z)\ a_-(z)\ -  \ f(z)\ a_-(z) \right)\  \nonumber \\
 &-&\left. 2\lambda\left(2z\left(r_{D^\star}^2+z-1\right)+r_\ell^2\left(4r_{D^\star}^2+z-4\right)\right)
\ f(z)\ a_+(z) \right]
\end{eqnarray}
where $\lambda$ in this expression is short for $\lambda(1,r_{D^\star }^2,z)$. This expression indicates that for this mode the new physics is not simply proportional to the SM because the contributions from the vector and axial-vector form factors enter with different signs in the term arising from $W$-$W^\prime$ mixing.

To judge the relative importance of the two terms: $W^\prime$ contribution vs $W-W^\prime$ mixing contribution, we provide a rough numerical estimate for their respective contributions to the decay rate:
\begin{eqnarray}
\Gamma(B\to D^\star \tau \nu) &=&  \frac{G_{\rm F}^2\,m_B^{ 5}}{192\pi^3} \  |V_{qb}|^2 \left(0.062 \ F^q_{W^\prime}\ - 0.11 \  F^q_{\rm Mix}    \right)
\end{eqnarray}
To obtain this estimate we converted the form factors to the usual HQET parameterization \cite{Manohar:2000dt} and used $\xi(w)=1$ for the Isgur-Wise function over the full $w=v_B\cdot v_{D^\star}$ range.

\section{Numerical Results and Conclusions}

The data on semileptonic $B$ decay into $\tau$ leptons Eqs.~\ref{babarsemi}, \ref{bellesemi} can be fit to our model in terms of the two combinations of constants defined in Eq.~\ref{newfs}. For example, if we use the central value of each experiment we obtain
\begin{eqnarray}
{\rm BaBar} &\Rightarrow & F^c_{W^\prime}=1.40 ,\,\, F^c_{{\rm Mix}}=0.043 \nonumber \\
{\rm Belle} &\Rightarrow & F^c_{W^\prime}=1.46 ,\,\, F^c_{{\rm Mix}}=-0.14
\end{eqnarray}
If instead we use the more precise BaBar result at the one sigma level we find
\begin{eqnarray}
 1.2 \lsim &F^c_{W^\prime}& \lsim 1.6 \nonumber \\
 0.009 \lsim & F^c_{{\rm Mix}} & \lsim 0.08
 \label{reqrange}
 \end{eqnarray}

To see whether these numbers can be produced in our model let us recall the existing constraints. 
The model described earlier is geared for an enhanced ratio of right-handed to left-handed couplings $g_R/g_L$. The constraints that LEP2 data impose on this ratio can be roughly summarized by the relation \cite{heval}
\begin{equation}
\frac{g_R}{g_L}\lsim \frac{M_{W^\prime}}{M_W}, 
\label{appbound}
\end{equation}
whereas the requirement that the new gauge interaction ($SU(2)_R$) remains perturbative implies that 
$g_R \lsim 10\ g_L$. 

The FCNC generated in this model have also been studied in detail and summarized in Ref.~\cite{He:2009ie}. There, an ansatz was produced to understand the smallness of the observed FCNC, and this ansatz gives us specific predictions for the quark-mixing angles that appear in Eq.~\ref{newfs}. Specifically, we found that 
\begin{eqnarray}
V^d_{Rbi} \sim \delta_{bi},\,\, V^u_{Rtc}\sim V_{cb}, \,\, V^u_{Rtu}\sim V_{ub} 
\end{eqnarray}
which then lead to
\begin{eqnarray}
V_{Rub}\sim V_{ub}  &&V_{Rcb}\sim V_{cb}.
\end{eqnarray}
Finally, as mentioned earlier, we will assume that the charged lepton mixing matrix is nearly diagonal so that $V^\ell_{R3\tau} \approx 1$. 
Under these conditions the fit of Eq.~\ref{reqrange}
requires 
\begin{eqnarray}
0.7 \lsim & \frac{g_RM_W}{g_L M_{W^\prime}} &\lsim  0.9 \nonumber \\
0.005 \lsim & \xi_W\frac{g_R}{g_L} &\lsim 0.06
\label{conditions}
\end{eqnarray}
The first of these conditions can be satisfied within the constraint Eq.~\ref{appbound} and the requirement that the new interaction remains perturbative provided $M_{W^\prime}\lsim 1$~TeV. 
This is a range that should be accessible to LHC, providing new motivation to search for a non-universal $W^\prime$.

The second condition requires that we re-examine bounds on $W-W^\prime$ mixing, in particular the combination $\xi_{eff}=\xi_W g_R/g_L$. This combination is constrained by $b\to s\gamma$~\cite{heval,Babu:1993hx} and we applied this constraint to our model in ref.~\cite{heval}. We can update that result using a recent HFAG average $B(b\to X_s \gamma) = (3.55 \pm 0.25)\times 10^{-4}$ \cite{Amhis:2012bh} for the experimental number and a recent NNLL SM calculation that gives $B(b\to X_s \gamma) = (3.15 \pm 0.23)\times 10^{-4}$ \cite{Misiak:2006zs}. Assuming that the new physics interferes constructively with the SM, at the 3$\sigma$ level the allowed range becomes 
\begin{equation}
-1.3 \times 10^{-3} \lsim \xi_{eff} \lsim 2.7 \times 10^{-3} 
\end{equation}
For destructive interference between the new physics and the SM there is a second, narrow, allowed range near  $\xi_{eff}=-0.028$.
We see that the constraint from $b\to s\gamma$  limits the effect of mixing to be somewhat lower than what is required for the fit Eq.~\ref{reqrange} at $1\sigma$.

Our conclusion from this study is that our model can accommodate an enhancement in the ratios $R(D)$ and $R(D^\star)$ of the order reported by BaBar and Belle (through $F^c_{W^\prime}$), but that it predicts approximately the same enhancement for both ratios once we incorporate the constraint from $b \to s \gamma$ on $F^c_{{\rm Mix}}$. 

Our model also leads to predictions for related modes. For the leptonic $B\to\tau \nu$ decay we find
\begin{eqnarray}
\frac{\Gamma(B^-\to \tau^- \nu)}{\Gamma(B^-\to \tau^- \nu_\tau)_{SM}} &=&F^u_{W^\prime}-2\ F^u_{{\rm Mix}}.
\end{eqnarray}
With the constraints discussed above, this implies that a number as large as implied by the BaBar data in Eq.~\ref{fbtau} is possible. However, the mixing angles involved are different, in this case 
$V_{Rub}$ instead of $V_{Rcb}$, so this enhancement is not necessarily equal to the one occurring for $R(D)$ and $R(D^\star)$ although it is expected to be of the same order.

The inclusive semileptonic decay $b \to c \tau^- \nu$ can also be enhanced 
\begin{eqnarray}
\frac{\Gamma(b \to c \tau^- \nu)}{\Gamma(b \to c \tau^- \nu)_{SM}} &=&F^c_{W^\prime}-0.8\ F^c_{{\rm Mix}}
\end{eqnarray}
and the constraints described above from $b\to s\gamma$  restrict this factor to be approximately the same as the one occurring in the exclusive modes.

Finally, the model also leads to a correlation between the enhancement observed in $R(D,D^\star)$ and the leptonic decay of $B^-_c$
\begin{eqnarray}
\frac{\Gamma(B_c^-\to \tau^- \nu)}{\Gamma(B_c^-\to \tau^- \nu_\tau)_{SM}} &=&F^c_{W^\prime}-2\ F^c_{{\rm Mix}} .
\end{eqnarray}

\begin{acknowledgments}

The work of X.G.H. was supported in part by NSC of ROC, NNSF(grant No:11175115) and Shanghai science and technology commission (grant No: 11DZ2260700) of PRC.
The work of G. V. was supported in part by DOE under contract number DE-FG02-01ER41155. We thank Jusak Tandean and Alper Hayreter for useful discussions. 

\end{acknowledgments}

\appendix

\end{document}